# Cryogenic Memory Architecture Integrating Spin Hall Effect based Magnetic Memory and Superconductive Cryotron Devices

Minh-Hai Nguyen[1][*], Guilhem J. Ribeill[1], Martin Gustafsson[1], Shengjie Shi[2], Sriharsha V. Aradhya[2][†], Andrew P. Wagner[1], Leonardo M. Ranzani[1], Lijun Zhu[2], Reza Baghdadi[3], Brenden Butters[3], Emily Toomey[3], Marco Colangelo[3], Patrick A. Truitt[4][‡], Amir Jafari-Salim[4][‡], David McAllister[4][§], Daniel Yohannes[4][‡], Sean R. Cheng[1][**], Rich Lazarus[1], Oleg Mukhanov[4][‡], Karl K. Berggren[3], Robert A. Buhrman[2], Graham E. Rowlands[1], Thomas A. Ohki[1]

[1]Raytheon BBN Technologies, Cambridge, MA 02138, USA

[2]Cornell University, Ithaca, NY 14850, USA

[3]Massachussetts Institute of Technology, Cambridge, MA 02139, USA

[4]HYPRES, Inc., Elmsford, NY 10523, USA

**One of the most challenging obstacles to realizing exascale computing is minimizing the energy consumption of L2 cache, main memory, and interconnects to that memory. For promising cryogenic computing schemes utilizing Josephson junction superconducting logic, this obstacle is exacerbated by the cryogenic system requirements that expose the technology's lack of high-density, high-speed and power-efficient memory. Here we demonstrate an array of cryogenic memory cells consisting of a non-volatile three-terminal magnetic tunnel junction element driven by the spin Hall effect, combined with a superconducting heater-cryotron bit-select element. The write energy of these memory elements is roughly 8 pJ with a bit-select element, designed to achieve a minimum overhead power consumption of about 30%. Individual magnetic memory cells measured at 4 K show reliable switching with write error rates below $10^{-6}$, and a 4x4 array can be fully addressed with bit select error rates of $10^{-6}$. This demonstration is a first step towards a full cryogenic memory architecture targeting energy and performance specifications appropriate for applications in superconducting high performance and quantum computing control systems, which require significant memory resources operating at 4 K.**

Since the invention of the integrated circuit, the exponential increase in computing predicted by Moore's law has made high performance computing (HPC) an essential driver for technological and scientific advances. With exascale computers on the horizon, it has become increasingly important to minimize the energy consumption of computation to meet the needs of HPC applications in a reasonable power envelope. However, the end of Dennard scaling and the difficulty of scaling down conventional CMOS technology made it important to look to alternative technologies that can continue to advance the performance of HPC systems. While many new computing paradigms have recently been proposed, such as in-memory[1,2] and neuromorphic[3,4] computing, it is still critical to improve the computational efficiency of conventional architectures.

Superconducting computers have long been a strong candidate for such "more than Moore" computing with the potential of two orders of magnitude lower power consumption than CMOS technology, even including the cost of cryogenic cooling power[1]. This large advantage is due to near-lossless signal propagation along superconducting wires, and the picosecond delay and atto-Joule switching energy of Josephson junctions (JJs). In recent years, Energy-efficient Rapid Single Flux Quantum[2] and Reciprocal Quantum Logic circuits[5] have demonstrated operation with zero static power dissipation. Practical implementations of superconducting logic are expected to operate at clock frequencies of up to 80 GHz[5,6], potentially an order of magnitude faster than their CMOS counterparts. Furthermore, the energy efficiency of superconducting electronics makes them an attractive technology for use as an interface to quantum computing systems that operate at millikelvin temperatures.

While superconducting logic has been extensively developed and demonstrated experimentally, fast and power-efficient cryogenic memory has been one of the main obstacles in realizing fully functional superconducting computers[1]. Memories based on single-flux-quantum (SFQ) technology[7] have large cell sizes and power dissipation[1]. CMOS-based memories[8–10] provide better scalability but also suffer from their high dynamic power consumption at cryogenic temperatures, in the range of 10-100 pJ/bit.

---

[*] Corresponding author. Email: minh-hai.nguyen@raytheon.com.
[†] Curren address: Western Digital Corp., Fremont, CA 94539, USA.
[‡] Current address: SeeQC Inc., Elmsford, NY 10523, USA.
[§] Current address: Bluemont Technology, Luray, VA 22835, USA.
[**] Current address: Harvard University, Cambridge, MA 02138, USA.

Although there is an active pursuit of other types of superconducting memories[11–14], a great deal of effort has been invested in magnetic[15–17] and magnetic-superconducting hybrid memories[18,19] which have the potential to offer high speed, low power, non-volatility, and scalability. Early achievements in cryogenic spin-valves[15,16] and magnetic tunnel junctions[17] (MTJs) have recently been reported with energy consumption per switching of a few tens of fJ. Integrating these successfully demonstrated magnetic elements with superconducting circuitry becomes an appealing route in the pursuit of larger scale cryogenic memory.

We report on the design and experimental characterization of a prototype cryogenic memory array comprising two key technologies: magnetic tunnel junctions based on the spin-Hall effect (SHE-MTJs)[20] as storage elements, and superconductive heater-cryotrons (hTrons)[11] as control elements, as illustrated in Fig. **1(a-c)**. The SHE-MTJs are three-terminal devices consisting of an MTJ atop a metallic spin Hall channel. Nanosecond magnetic reversal time and write error rates < $10^{-5}$ for writing current densities as low as $8\times10^{11}$ A/$m^2$ at room temperature (RT) have been reported for this class of device[21]. Our recent work confirmed similar performance at 4 K[17] providing a high-speed and low-power solution for cryogenic memory. hTrons are a variant of the nano-cryotron (nTron)[22] logic family, which can be driven by SFQ logic and support sufficient current to switch the magnetic element.

## ARCHITECTURE DESCRIPTION

The cryogenic memory system we developed includes a memory array and superconducting control circuits. The latter are currently fabricated on separate chips which can be connected to the memory array by wire-bonding or Multi-Chip Module (MCM) technology. The control circuits, designed with SFQ technologies, can supply triggering signals of ~100 μA into a load of a few Ohms. Here we limit our discussion to the design and performance of the main memory array which will target compatibility with the specifications of these control chips.

### *Memory Array Architecture*

As illustrated in Fig. **1(a-c)**, our proposed memory cell consists of a SHE-MTJ device in parallel with a bit-select hTron device. The addressing architecture was based on a design described in Ref. 23. The SHE-MTJ, as depicted in Fig. **1(a)**, is a three-terminal device consisting of an MTJ patterned on top of a spin Hall channel. The MTJ itself is an elliptical nanopillar consisting of two ferromagnetic layers separated by a thin tunnel barrier: the reference layer's (RL) magnetization is rigidly pinned, while the free layer's (FL) magnetization can be switched parallel (P) or antiparallel (AP) to that of RL. One bit of information is encoded in this non-volatile relative orientation: the tunneling magnetoresistance (TMR) effect gives rise to two easily distinguishable resistance states ($R_P$ or $R_{AP}$) in either case. In our measurements, memory readout is performed by monitoring the MTJ voltage $V_{MTJ}$ while applying a small 5 μA sense current through the MTJ.

To switch the magnetization of the FL, i.e., to change the MTJ state, a current is applied in the spin Hall channel. The SHE[24–26] of the channel induces spin accumulations on the surfaces of the channel, in turn enacting magnetization reversal of the FL via the spin (transfer) torque[20,27,28] as spins diffuse into the ferromagnet. The low switching current and confinement of the spin-torque switching mechanism within the device (no external magnetic fields, with their detrimental impact on superconductive circuits, are required) make the SHE-MTJ suitable for integration with superconducting structures.

An issue with the SHE-MTJ, however, is that its characteristic impedance and switching currents are too large to be directly compatible with our separately fabricated SFQ control circuits. For example, a 300 nm wide, 5 nm thick spin Hall channel requires a switching current of roughly 1 mA into a 0.5 kΩ load, which is incompatible with typical SFQ circuit output impedance of a few Ohms.

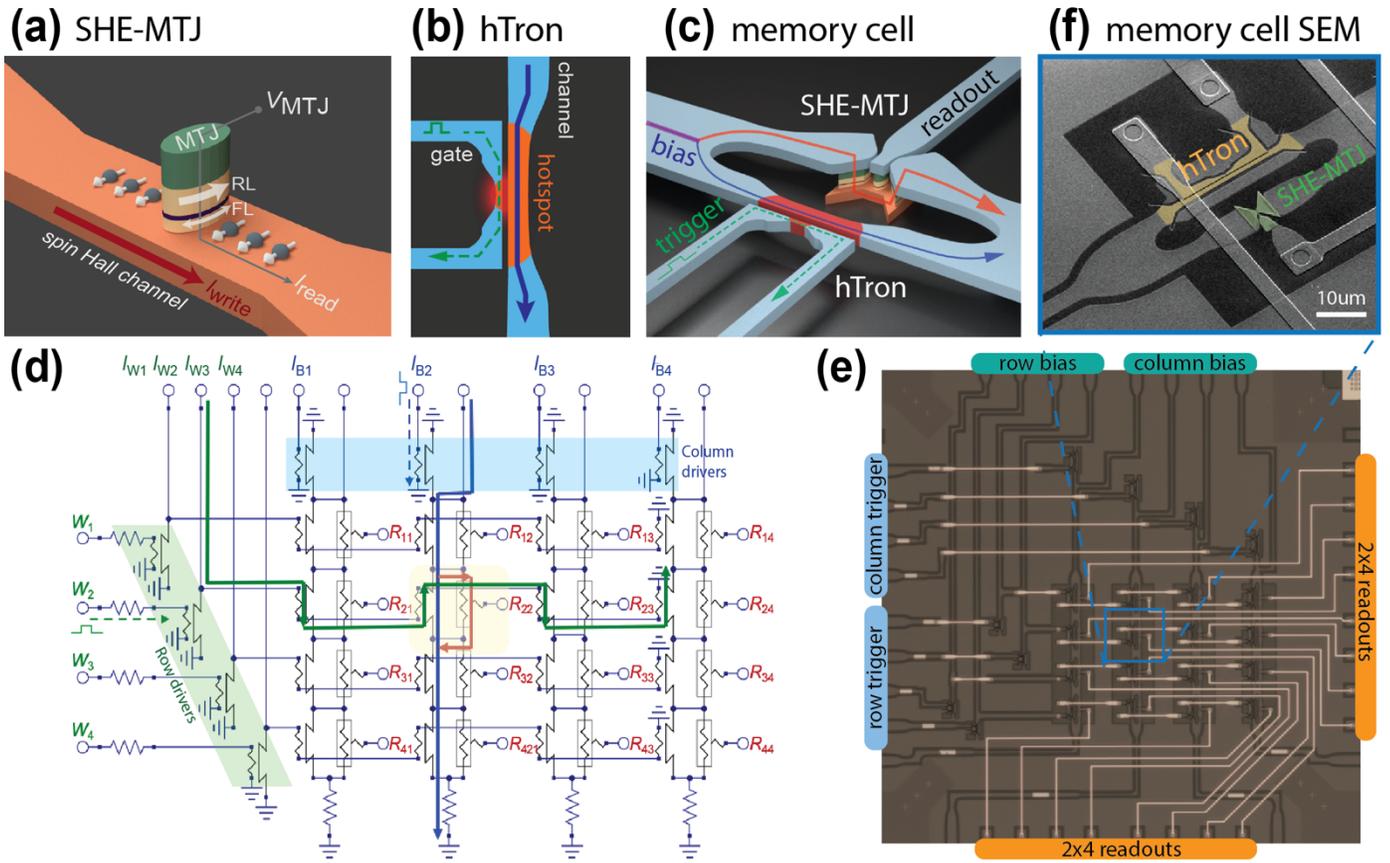

**Figure 1. Proposed cryogenic memory architecture.** Not-to-scale illustrations of **(a)** a SHE-MTJ device, **(b)** a hTron device and **(c)** a memory cell. **(d)** Schematic of the 4x4 memory array interfaced with external control circuitries by column and row drivers. The arrows indicate the current paths during the writing process of cell (2,2). **(e)** Micrograph of the memory array. **(f)** SEM image of cell (2,2) in the array.

To resolve this problem, we use a heater-cryotron (hTron) bit-select element connected in parallel with the SHE-MTJ channel which acts as a three terminal switch and also can be engineered to have better impedance match to the SHE-MTJ. The hTron[11] is a non-contact variation of the nano-cryotron[22] in which heat generated in a gate nanowire to suppress the critical current of an adjacent superconducting channel below its nominal operating current, thereby switching it into the normal state. Our implementation uses a nanowire gate constriction made from the same material and placed in the same plane and very near to the main channel, as illustrated in Fig. **1(b)**. A small triggering pulse (green arrow) is applied to the gate constriction, sending it into the normal state and creating a hot spot because of dissipative current flow. The lateral heat flowing away from this hotspot suppresses the critical current of the nearby channel and causes the channel to enter the normal state if it is biased (blue arrow). The resistance of the channel in the normal state can be engineered to reach multiple kilo-Ohms while carrying in excess of 1 mA. Our primitive memory cell is formed by placing the hTron in parallel with a SHE-MTJ channel, as depicted in Fig. **1(c)**. Triggering the hTron presents a large enough impedance that current redirects through the SHE channel and switches the MTJ's state. In this manner, the memory element can be controlled by a small hTron gate pulse (100 µA) sourced by a line-driver hTron which can be interfaced directly with the SFQ peripheral circuitry. The use of an hTron or an equivalently high isolation device is particularly important in the interior of the memory arrays: sneak currents or leakage current from writing to a cell can otherwise cause redistribution of current within the surrounding circuitry, since in the inactive state all lines are superconducting. This can create multiple current loops for leakage current to accumulate. This effect is mitigated by inclusion of a bit select element that has high isolation. In addition, when an hTron is inactivate, current simply flows through the superconducting path in the memory cell to subsequent cells in the columns and is recycled elsewhere with no concerns of other leakage paths.

Scaling to a device array is accomplished by chaining these memory cells into many parallel columns. The schematic of a 4x4 memory array is shown in Fig. **1(d)**. To be able to address the individual memory cells, the array is connected to peripheral row and column drivers, which are also hTron devices, to address the individual memory cells. The writing procedure is as follows: All row and column driver hTrons are biased with currents, which are initially shunted to ground through the driver channels. To write to a cell, a triggering pulse is applied to the corresponding column driver hTron (dashed blue arrow applied at $I_{B2}$). The column driver becomes normal and consequently the bias current is diverted down the column of cells and flows through all bit-select hTrons in the column (solid blue arrow), which are still superconducting. Any combination of columns can be activated, with each column constituting a bit in the 4-bit word to be written. Next, one row driver hTron is triggered (dashed green arrow), diverting its bias current through the gates of all bit-select hTrons in that row (solid green arrow). For those bit-select hTrons which have both the current from the column driver flowing through their channel and receive a gate current from the row driver, the triggering condition is fulfilled, diverting sufficient current to the SHE-MTJ channel to produce a change in magnetic state (orange arrow).

The micrograph of the 4x4 memory array is shown in Fig. **1(e)**. A scanning electron microscopy (SEM) image of cell (2,2) in the array is shown in Fig. **1(f)**. See Methods for detail on sample multilayer and fabrication process.

*Optimizing for Power Efficiency*

Because current is shared between the channels of the SHE-MTJ and hTron bit-select elements in the memory cell during writing events, it is important to match their impedances to minimize power consumption. For fixed SHE-MTJ parameters, the problem becomes determining an hTron design which minimizes dissipation while still providing enough current to switch the MTJ. This means that the normal-state resistance of the hTron should be maximized, so that most of the current will be diverted into the memory cell, with the following two constraints:

(i) The column bias current must be lower than the hTron critical current of the hTron channel, to ensure that the bit-select hTron channel is superconducting before its gate is triggered.

(ii) The current through the hTron channel during the writing event must be large enough to sustain the hotspot in the channel[29].

Solving straightforward inequalities (see Methods), the optimal width and length of the hTron are found to be:

$$W_0 = \frac{I_{SH}}{J^c d - \sqrt{2\alpha\Delta T/R_{sq}}}, \quad (1)$$

$$L_0 = \frac{I_{SH} \cdot R_{SH}}{\sqrt{2\alpha\Delta T R_{sq}}}, \quad (2)$$

where $I_{SH}$ is the SHE-MTJ switching current, $R_{SH}$ its channel resistance; $J^c$, $d$, $R_{sq}$, $\alpha$ are hTron channel's critical current density, thickness, sheet resistance, and specific heat per unit area, respectively; and $\Delta T = T_c - T_s$ is the difference between the hTron critical temperature $T_c$ and the sample temperature $T_s$.

In this optimal design, the overhead power consumption added by the hTron is then (see Methods):

$$P_{\text{overhead}} = \frac{\sqrt{2\alpha\Delta T/R_{sq}}}{J^c d - \sqrt{2\alpha\Delta T/R_{sq}}}, \quad (3)$$

which is fixed by the hTron's material properties and system temperature, independent of SHE-MTJ parameters. This result implies that we can always design the bit-select hTron to accommodate the impedance of a particular SHE-MTJ device with a fixed overhead power. Using typical values for our NbN film (not fully optimized for this application), $\alpha = 220$ kW/m$^2$K, $R_{sq} = 100$-$140$ Ω/sq, $T_c = 12.8$ K, $T_s = 3.6$ K, $J_c = 25$-$30$ GA/m$^2$, the minimal energy overhead $P_{\text{overhead}}$ is 30(6)%. Estimated using the results from nanosecond pulse switching of a standalone SHE-MTJ device, described in the next section, the typical switching energy for a SHE-MTJ device is roughly 6 pJ and thus that of a memory cell is about 8 pJ per switching.

**MEASUREMENTS AND RESULTS**

*Standalone SHE-MTJ device*

As described above, our memory element, the spin-Hall-effect based magnetic tunnel junction (SHE-MTJ), consists of an MTJ atop a spin Hall channel, as illustrated in Fig. **1(a)**. A writing (electrical) current in the channel induces a transverse spin accumulation that exerts torque on the free layer (FL) via the SHE, thereby switching its magnetization back and forth. The conventional figure of merit representing the strength of the SHE is the spin Hall angle, which is the ratio of the induced spin current to the applied electrical current. Energy-efficient switching devices require metallic materials possessing very strong SHE such as Pt[30], β-Ta[20], β-W[31] and their alloys[32–34] whose (absolute) spin Hall angles are in the range of 0.1-0.35.

SHE-MTJ devices in our memory arrays employ a Pt$_{85}$Hf$_{15}$ alloy channel with spin Hall angle ~0.2. This was an early design decision made with fabrication stability in mind, and without the benefit of subsequent materials advances. As seen in the SHE-MTJ stack illustrated in Fig. **2(a)**, a thin Hf layer of 0.5 nm thick is inserted between the channel and the MTJ to suppress the Gilbert magnetic damping caused by spin pumping[35]. Further device details are given in the Methods section.

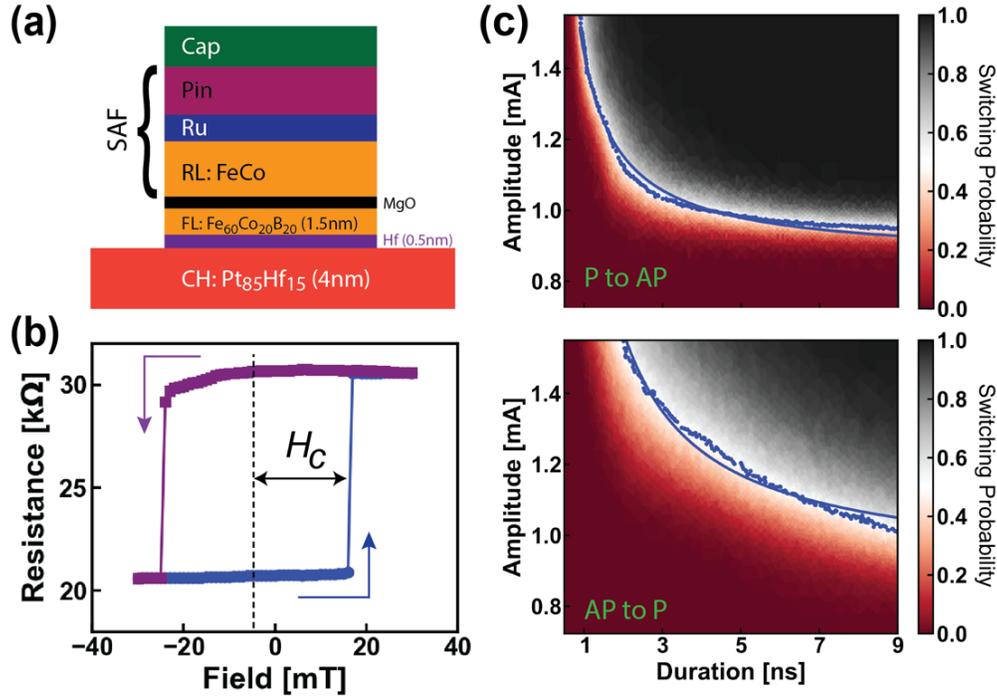

**Figure 2. Cryogenic characterization of a standalone SHE-MTJ device. (a)** SHE-MTJ multilayer consisting of a spin Hall channel, Hf spacer and MTJ stack pinned by a synthetic antiferromagnetic (SAF) structure. **(b)** Magnetic minor loop of the MTJ with a 3 µA sense current. The dotted line shows the center of the loop (stray field). **(c)** Switching probability (encoded by the color) as a function of pulse amplitude and duration for P to AP (top) and AP to P (bottom). Blue dots are points at 50% switching probability which are fitted by the macrospin model (blue curves).

We find robust SHE-MTJs response seen in other devices of similar composition[36]. The tunneling magnetoresistance (TMR) response to an easy-axis magnetic field is shown in Fig. **2(b)**. The MTJ switches between the parallel (P, low resistance) and antiparallel (AP, high resistance) states with a 20 mT coercivity (loop half-width) and a small offset field of -5 mT. Bi-stability at zero external field is thus achieved: a mandatory criterion given the flux-trapping propensity of the superconducting peripheral circuitry. Another criterion is satisfied in that the AP and P resistance states are easily discernible using a simple superconducting comparator.

The nanosecond pulse switching performance of the SHE-MTJ device is gauged using the same measurement setup and technique outlined in our previous reports[17,36]. The switching probability is shown in Fig. **2(c)** for the two switching polarities as a function of pulse duration and amplitude. We fit the pulse amplitudes $I_{50\%}$ and pulse durations $t_{50\%}$ along the 50% probability boundary (blue dots) with the macrospin model for ballistic reversal (blue curves)[37,38] to obtain the critical current $I_0$ and characteristic reversal time $t_0$. For P to AP switching we find $I_0$=0.9 mA and $t_0$=0.7 ns while for AP to P switching we find $I_0$=0.9 mA and $t_0$=1.5 ns. These $t_0$ values are similar to previous measurements (at 4K) on Pt channel devices[17], while the $I_0$ values are about two times lower, which is consistent with the two times higher spin Hall angle of $Pt_{85}Hf_{15}$ when compared to pure Pt[36]. With the performance of the memory element itself confirmed, we verify the ability of our hTron select element to redirect its channel current across the SHE-MTJ to actuate switching.

### Standalone hTron device

The heater cryotron (hTron) devices shown in Fig. **1(b)** are arranged in parallel with a SHE-MTJ devices as shown in Fig. **1(c)**. A current pulse of 100 µA is supplied the hTron gate, ultimately sourced by a SFQ-DC or SFQ-RO (relaxation oscillator) driver, sending the gate into the normal state and producing sufficient Ohmic heating to trigger hotspot formation in the hTron's current carrying channel situated 100 nm away. The supercurrent is impeded by the emergence of a ~1kΩ dissipating impedance of the normal state NbN, and a large fraction of the current is redirected into the SHE-MTJ channel where it switches the memory element. To minimize the overhead power consumption of the bit-select hTron (see analysis in the previous section), the hTron channel was designed to be 1.7 µm × 25.5 µm, which gives the optimal overhead energy consumption about 31%.

Before studying integrated devices, we performed quasi-static switching measurements on one standalone hTron device. As shown in Fig. **3(a)**, we biased the hTron channel with a square pulse (blue) and the gate with a triangular pulse (red), while monitoring their respective voltages (right axis). The voltages remain at the noise floor until the gate current reaches the critical value of 87 µA, beyond which the gate becomes normal (resistive). If the channel current is high enough, as is the case in Fig. **3(a)**, the channel also becomes normal. Beyond $t = 0.3$ ms, the resistive state of the channel is maintained by the channel current (below its critical value) even though the gate current has been turned off. This important latching behavior is a result of thermal run-away in the channel as its resistance increases and allows the triggering pulse to the gate to be short and of low amplitude.

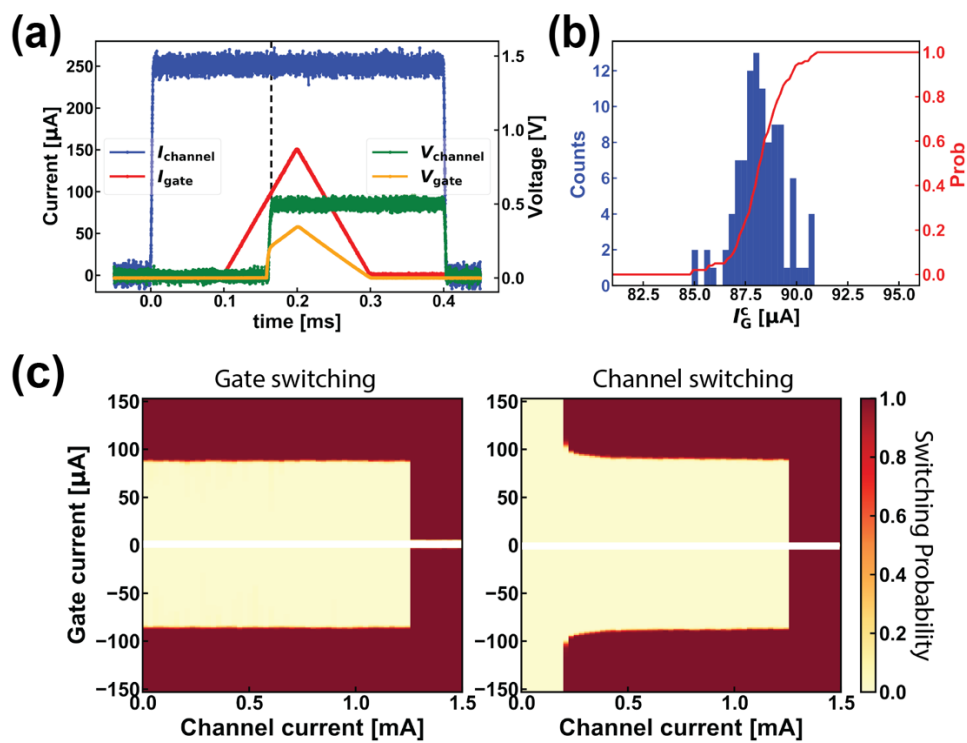

**Figure 3. Characterisation of a standalone hTron device.** (**a**) Voltage drops (right axis) over the hTron gate and channel as functions of applied currents (left axis). Switching occurs at $I_G^c = 87$ µA. (**b**) Distribution of the gate critical current obtained by repeating the bias sequence in panel (**a**) 100 times, and its cumulative probability of switching to the normal state (right axis). (**c**) Probabilities for the gate (left) and channel (right) of switching to the normal state as functions of applied gate and channel currents.

We repeated the measurement in Fig. **3(a)** 100 times and measured the distribution of current required to turn the gate resistive, shown in Fig. **3(b)** along with the cumulative switching probability. We acquired repeated histograms of this type, for a range of channel currents, and the resulting switching probability is shown in the left panel of Fig. **3(c)**. The clear and orthogonal switching boundaries shows that the gate and channel of the hTron are galvanically isolated, as expected.

We next performed a similar experiment but this time monitored the switching (to normal state) of the hTron channel, and obtained the results shown in the right panel of Fig. **3(c)**. Comparing the two panels of Fig. **3(c)** reveals an interesting correlation: For $|I_{gate}| > 90$ µA, the gate always turns normal, but the channel state does not necessarily follow. Instead, the boundaries around $I_{channel} = 0.2$ mA and $|I_{gate}| = 90$ µA are curved. This difference from the orthogonal boundaries seen in the left panel is due to the relative sizes of the hotspots created in the channel and the gate.

*Memory cell*

To perform switching experiments on a memory cell — as shown in Fig. **1(c)** — we used external pulse generators to provide bias and triggering current pulses via reference resistances of 5-10 kΩ, as shown in Fig. **4(a)**. To measure the MTJ resistance, we applied a DC bias current of 5 µA into the top of the MTJ and monitored its voltage drop.

We performed microsecond pulse switching measurement of the representative memory cell (2,2) in the array at 4 K without an external magnetic field (see Figure **1(d)**). In these experiments, we biased the 2nd row driver hTron with a 1 mA pulse and applied a trigger pulse of 100 μA to the gate. The gate trigger level of the 2nd column driver hTron was also 100 μA, but its channel bias was varied between -1.5 and 1.5 mA. Figure **4(b)** shows the traces of the input and sample voltages read from an oscilloscope (for column bias = -1.5 mA). The MTJ state is determined by the sign of $\Delta V_{MTJ}$ which is the MTJ drop voltage subtracted by its median value of 53 mV.

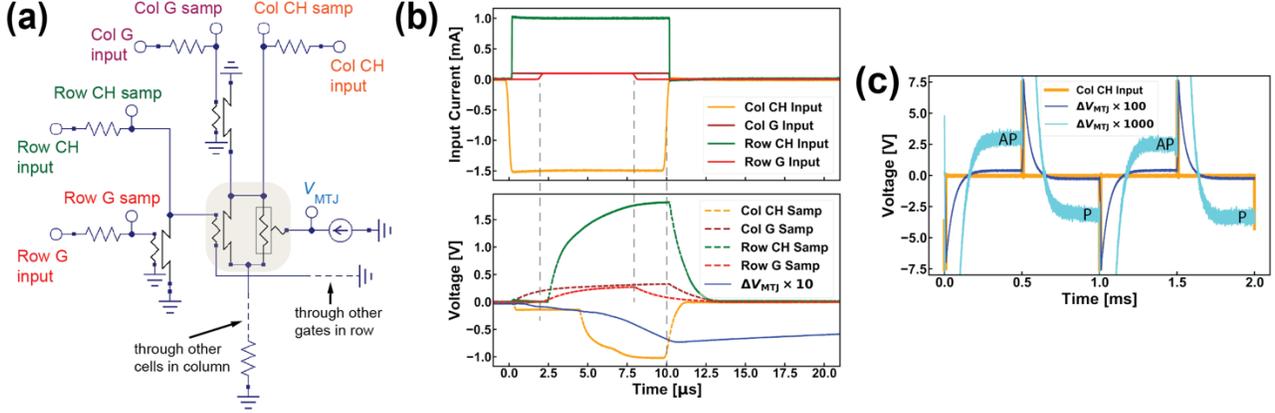

**Figure 4. Switching of memory cell (2,2). (a)** Schematic of the measurement circuitry on a representative cell (1,1) in the array (other cells not shown). **(b)** Input (top) and sample (bottom) voltage traces of column and row channel (CH) and gate (G). Blue trace (bottom) shows the variation of the MTJ voltage $\Delta V_{MTJ}$. **(c)** Column bias and enlarged $\Delta V_{MTJ}$ show the MTJ state's responding to the sign of the column bias.

As shown in Fig. **4(b)**, for time $t < 2$ μs, the column bias current flows through the superconductive hTron channels of the 2nd column and a small sensing resistance at the bottom, resulting in a small voltage drop. At $t = 2$ μs, the row driver is triggered, and after a 2.5 μs delay (due to the length of the coaxial cables of a few meters from the chip package in the cryostat to the RT electronics) at $t = 4.5$ μs the row driver channel builds up enough current to trigger the bit-select hTron, resulting in a rapid rise in the column voltage. This result confirms the intended operation of the cell and driver hTrons.

The response of $\Delta V_{MTJ}$ to the column bias is shown in Fig. **4(c)**. When the 6 μs long column bias current (orange line) is negative, the triggering of the bit-select hTron causes the MTJ voltage to relax to the AP state ($\Delta V_{MTJ} > 0$). Conversely, when the column bias current is positive, the MTJ voltage relaxes to its P state ($\Delta V_{MTJ} < 0$). This clearly demonstrates the successful control of the memory cell (2,2) in the array. We note that the long rise and fall times observed in Figs. **4(b,c)** result from the use of slow external electronics, and are not characteristic of the underlying devices.

We determined the write-error-rate (WER) of the cell by repeating the switching experiment $10^3$ - $10^6$ times, with each attempt preceded by resetting the MTJ to its original state by applying a high current bias to the column. The resulting behavior, shown in Fig. **5(a)**, is very similar in shape to the previously reported WER of single standalone SHE-MTJs[17,21]. Note that the open symbols near the bottom of the plot indicate no write error was observed in 1 million switching attempts, implying that the WER is below $10^{-6}$ for bias current of 1.0 mA (P to AP) and 1.5 mA (AP to P).

We characterized the intercell crosstalk by monitoring the MTJ voltage of cell (2,2) while triggering its four neighboring cells. Fig. **5(b)** shows the response in $\Delta V_{MTJ}$ of cell (2,2) when the adjacent cell (2,1) was triggered. A voltage rise of about 3 mV is observed during triggering events, which translates to roughly 10 μA of leakage current in the SHE-MTJ channel of cell (2,2). This leakage current is too small to affect the state of the cell (2,2), which remains unchanged in the AP state. We repeated the experiment with all permutations of cell (2,2) original state, bias current polarity, switching each of the four neighboring cells, with $10^4$ switching attempts for each case. No unintended switching of cell (2,2) was observed.

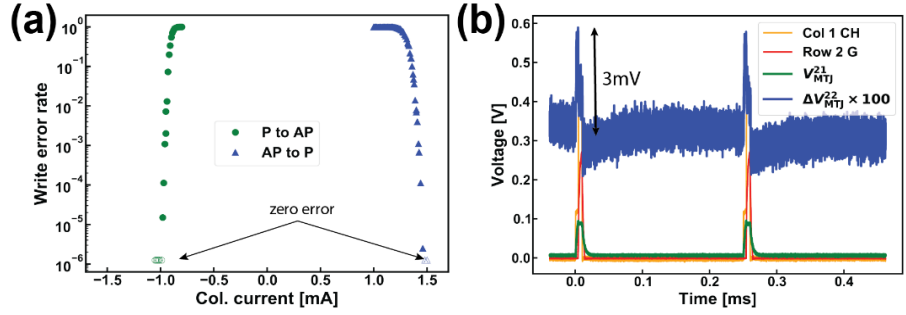

**Figure 5. Write error rate and crosstalk of memory cell (2,2).** (**a**) Memory write error rate (log scale) as a function of column bias current. The open symbols indicate zero error observed in $10^6$ switching attempts. (**b**) Enlarged $\Delta V_{MTJ}$ of cell (2,2) when triggering cell (2,1) shows a leakage voltage of 3 mV.

*Array triggering fidelity*

We then investigated the triggering fidelity (successful rate) of the 4x4 memory array by writing all possible 4-bit words to each row and detecting whether the targeted bit-select hTrons triggered. the measurement procedure is as follows: The column and row driver hTron channels are biased with long current pulses. For each bit in the word to be written, the corresponding column driver is activated with a 1 µs long pulse on its gate. The voltage at the top of each column, $V_{pre}$, is measured, to make sure that none of the bit-select hTrons have been accidentally activated before the row driver. One of the four row drivers is then activated with a 1µs long current pulse, and the column voltages $V_{1-4}$ are measured again. The write is considered successful if the following criteria are fulfilled:

(i) All column voltages fall below a threshold before the row is triggered: $V_{pre} < V_{min}$.

(ii) All column voltages $V_{1-4}$ for unwritten bits fall below a threshold value after the row is triggered: $V_{1-4}$ (unwriten) $< V_{min}$.

(iii) All column voltages $V_{1-4}$ for written bits fall within a specified interval after the row is triggered: $V_{min} < V_{1-4}$ (written) $< V_{max}$.

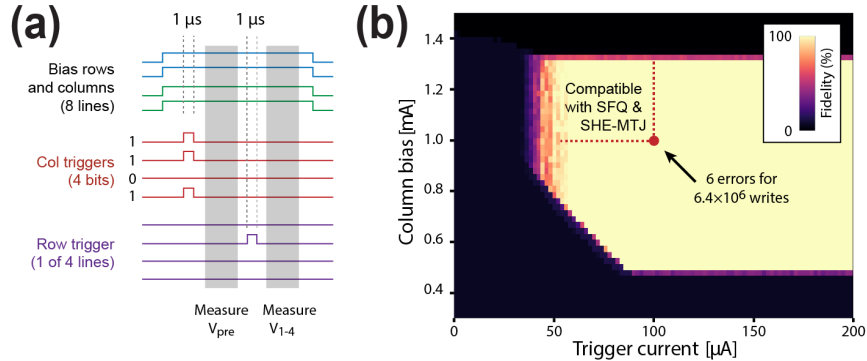

**Figure 6. Fidelity of the array triggering.** (**a**) Illustration of a pulse sequence of the fidelity measurements. The column voltages are measured before and after row triggering. (**b**) Triggering fidelity versus column bias current and trigger current with 640 writing sequences for each pixel. The red dotted lines indicate the constraints set by the memory element and SFQ control circuits.

Figure **6(b)** shows the measured triggering fidelity with respect to column bias current and the trigger current used to activate both row and column drivers. The fidelity is assessed from 640 written words per pixel, encompassing all 16 possible bit patterns, written ten times to each row. We found that the fidelity is nearly 100% for a wide range of parameter values. Here we used $V_{min}$ = 400 mV, $V_{max}$ = 1 V, and a row bias current of 0.8 mA, but found that the measured fidelity did not depend critically on any of these values.

As previously determined, a column current of ~1 mA suffices to reliably change the state of a SHE-MTJ element, and SFQ circuitry is able to provide trigger pulses of ~100 μA. For this critical combination of parameters, highlighted with a red spot in Fig. **6(b)**, we benchmarked the writing fidelity in detail with 6.4 million writing cycles (distributed over all rows and bit patterns as above) and yielded a total of 6 errors. The bright area confined by the dotted line in Fig. **6(b)** indicates the parameter space suitable for the operation of the SHE-MTJ elements and SFQ control pulses.

**CONCLUSION AND OUTLOOK**

We have experimentally demonstrated the successful integration of SHE-based MRAM technology and the superconductive hTron bit-selects and drivers in a 4x4 memory array architecture which can be triggered by current signals as low as 100 μA, compatible with SFQ DC/RO control circuits. The array is fully addressed with an error rate as low as $1 \times 10^{-6}$, while the write-error-rate of the individual memory cells is shown to be below $10^{-6}$. The bit-select hTron was designed to match the impedance of the SHE-MTJ channel, adding a minimal energy overhead of 31% to the switching energy which is currently in the order of 6 pJ per switch.

Beyond the proof-of-concept demonstration of the cryogenic memory array architecture presented in this Report, we anticipate to make significant improvements to array speed and energy consumption. Because the minimum energy overhead added by the bit-select hTron is indpendent of SHE-MTJ parameters, the overall energy consumption of the memory cell scales with that of the SHE-MTJ element which can be further decreased using newly discovered spin Hall materials[34,39], device structures[40] and better lithography technologies. For example, the SHE-MTJ channel can be replaced with materials such as $Au_{1-x}Pt_x$ alloys whose spin Hall angles approach 0.35[34] for a four-fold energy reduction. The SHE-MTJ channel resistance, currently about 0.6 kΩ, whose major contribution comes from the two vias providing electrical contact to the SHE channel (see Fig. **1(c)**) can be reduced 2-3 times by replacing them with metallic vias. Another obvious improvement in design is scaling down the SHE-MTJ channel to reduce the switching current. Combining all these modifications in materials and design, we estimate the cell's switching energies could be as low as 0.1 pJ, more than 2,000 times less than that of cryogenic DRAM[9].

There are also opportunities to improve the performance of the control elements. The hTron column and row drivers used in our prototype array, whose triggering delay is roughly 8 ns[11] due to the heat propagation across the 100 nm gap in the film plane, can be replaced with nTrons[22] with vanishing triggering delay owing to the galvanic connectivity of their gate and channel. Our hTron bit-selects can also be replaced with a stacked variant[14] in which the gate crosses over the channel thereby providing more efficient heating and smaller triggering current. These enhancements will result in a faster, more energy efficient, and more compact bit-select and driver elements.

By combining the high speed and non-volatility of the MRAM, the superconductivity of the select elements, and the compatibility with SFQ circuitry and fabrication, our cryogenic memory architecture is proven to be a fast and power efficient candidate for exascale superconducting computing and quantum control systems.

**METHODS**

**Samples:**

As illustrated in Fig. **2(a)**, the SHE-MTJ stack is as follows (thicknesses in nanometer): *substrate* / $Pt_{85}Hf_{15}(4)$ / $Hf(0.5)$ / FL / MgO / RL / Ru / PL / cap, where FL = 1.5 nm $Fe_{60}Co_{20}B_{20}$, RL is the magnetic fixed layer pinned by the pinning layer PL in the synthetic antiferromagnetic structure RL / Ru / PL which is tuned to minimize the stray dipole field at the FL so that the magnetization of the FL is bistable at zero bias field. The stack was patterned into SHE-MTJ devices with 300 nm wide channel and 75 nm × 190 nm nanopillar MTJ by deep-UV and e-beam lithography and Argon ion milling.

Upon the fabrication of SHE-MTJ devices, a NbN layer of 30 nm thick was deposited at ambient temperature and patterned by photo- and electron-beam lithography. Superconducting structures are designed using conformal mapping curves[41] to prevent current crowding effect at corners. The bit-select hTron channel size was 1.7 μm × 25.5 μm, which was optimized to achieve the minimal energy overhead of 31%. A second level of metallic interconnect was provided by e-beam evaporated aluminum wires and vias and PECVD SiN dielectric. Micrographs of the array and a representative cell are shows in Figs. **1(e-f)**.

**Impedance matching**:

To minimize the overhead power consumption added to the SHE-MTJ switching by the bit-select hTron, the normal-state resistance of the hTron should be maximized, within the following two constraints:

(i) The column bias current must be lower than the hTron critical current of the hTron channel: $I_{bias} < J^c W d$ where $J^c$, $W$, $d$ are the critical current density, width and thickness of the hTron channel.

(ii) The current through the hTron channel during the writing event must be large enough to maintain the hotspot in the channel[24]: $J^2 d^2 R_{sq} > 2\alpha\Delta T$ where $J$ is the current density in the hTron channel during writing, $R_{sq}$ and $\alpha$ are its sheet resistance and specific heat per unit area in normal state, and $\Delta T = T_c - T_s$ is the difference between the hTron critical temperature $T_c$ and the sample temperature $T_s$.

Combining the above two conditions and the relation where $I_{SH}$ is the SHE-MTJ switching current, we obtain:

$$J^c W d > \sqrt{2\alpha\Delta T/R_{sq}} \cdot W + I_{SH}. \quad (M1)$$

Solving for $W$, we get

$$W > \frac{I_{SH}}{J^c d - \sqrt{2\alpha\Delta T/R_{sq}}} \quad (M2)$$

which implies equation (1) in the main text. Upon determining the minimum width, to find the optimal hTron channel length $L$, we use the condition (ii) and notice that $JWd = I_{SH} R_{SH}/(R_{sq} L/W)$ where $R_{SH}$ is the SHE-MTJ channel resistance, then obtain

$$L < \frac{I_{SH} \cdot R_{SH}}{\sqrt{2\alpha\Delta T R_{sq}}} \quad (M3)$$

which gives equation (2). The overhead power consumption added by the hTron is then (using (ii) and (M2)):

$$P_{\text{overhead}} = \frac{I_{\text{bias}}}{I_{SH}} - 1 = \frac{JWd}{I_{SH}} > \frac{W\sqrt{2\alpha\Delta T/R_{sq}}}{W(J^c d - \sqrt{2\alpha\Delta T/R_{sq}})} \quad (M4)$$

which yields equation (3), independent of SHE-MTJ parameters.

**Measurements:**

All measurements presented in this Report were performed at the base temperature of 3.6 K in cryogen-free cryostats. Characterizations of stand-alone SHE-MTJ devices (shown in Fig. **2**) were done using the same HPD cryostat and control electronics as described in our previous report[13]. Measurements and data collection procedures were performed using our custom-built Python package, Auspex, available online at *https://github.com/BBN-Q/Auspex*.

ACKNOWLEDGEMENTS

The authors thank B. Hassick and A. Kreider of Raytheon BBN; M. Onen, I. Charaev, and A. McCaughan of MIT for technical assistance; J. Walter, M. Kamkar, A. D'Addario, I. Vernik, J. Vivalda and M. Renzullo of HYPRES; and Synopsys, Inc. for technical assistance. The magnetic films for our SHE-MTJ devices were deposited by Canon ANELVA, Inc. The majority of nanofabrication in this work was performed at Cornell NanoScale Science and Technology Facility (NSF Grant NNCI-1542081) and Harvard Center for Nanoscale Systems (NSF Grant No. 1541959), members of National Nanotechnology Coordinated Infrastructure Network.

The research is based upon work supported by the Office of the Director of National Intelligence (ODNI), Intelligence Advanced Research Projects Activity (IARPA), under the Cryogenic Computing Complexity program[42] via contract W911NF-14-C0089. The views and conclusions contained herein are those of the authors and should not be interpreted as necessarily representing the official policies or endorsements, either expressed or implied, of the ODNI, IARPA, or the U.S. Government. This document does not contain technology or technical data controlled under either the U.S. International Traffic in Arms Regulations or the U.S. Export Administration Regulations.

**AUTHOR CONTRIBUTIONS**

Supervision: R.L., G.E.R., T.A.O., R.A.B., O.M., and K.K.B.

Device and chip design: M.-H.N., G.J.R., M.G., A.P.W., L.M.R., G.E.R., P.A.T., A.J.-S., D.M., and B.B.

Fabrication: G.J.R., A.P.W., S.S., S.V.A., L.Z., P.A.T., D.Y., E.T., M.C., and R.B.

Measurements and analyses: M.-H.N., M.G., S.R.C., S.S., A.J.-S., R.B., and B.B.

Manuscript composition: All authors.

**Competing financial interests**: The authors declare no competing interests.